\documentstyle[12pt,epsf]{article}

\advance\voffset by -1.5cm
\advance\hoffset by -1.5cm
\def\be{\begin{equation}}
\def\ee{\end{equation}}

\def\Zop{\bbbz}

\def\pmb#1{\setbox0=\hbox{#1}
 \kern-.025em\copy0\kern-\wd0
 \kern.05em\copy0\kern-\wd0
 \kern-.025em\raise.0433em\box0 }

\def\3{\ss}
\def\sq{\hbox{\rlap{$\sqcap$}$\sqcup$}}
\def\qed{\ifmmode\sq\else{\unskip\nobreak\hfil
\penalty50\hskip1em\null\nobreak\hfil\sq
\parfillskip=0pt\finalhyphendemerits=0\endgraf}\fi}
\def\half {\frac{1}{2}}

\def\bbbz {{\sf Z\!\!Z}}
\def\bbbr {{\rm I\!R}}

\def\Tr{{\rm Tr}}
\def\det{{\rm det}}
\def\R{{\cal R}}

\def\ss{\bf S}
\def\vv{\bf V}
\def\aa{\bf A}
\def\ii{\bf I}

\begin{document}

\thispagestyle{empty}
\def\thefootnote{\fnsymbol{footnote}}
\begin{flushright}
  hep-th/9711098\\
  HUTP-97/A048 \\
  DAMTP-97-119 \\
  PUPT-1730
\end{flushright}
\vskip 0.5cm

\begin{center}\LARGE
{\bf String Creation and Heterotic--Type I' Duality}
\end{center}
\vskip 0.5cm
\begin{center}\large
       Oren Bergman
       $^{a}$\footnote{E-mail  address: {\tt
bergman@string.harvard.edu}},
       Matthias R. Gaberdiel
       $^{b}$\footnote{E-mail  address: {\tt
M.R.Gaberdiel@damtp.cam.ac.uk}} and
       Gilad Lifschytz
       $^{c}$\footnote{E-mail  address: {\tt
       gilad@puhep1.princeton.edu}}
\end{center}
\vskip0.5cm

\begin{center}\it$^a$
Lyman Laboratory of Physics\\
Harvard University\\
Cambridge, MA 02138
\end{center}

\begin{center}\it$^b$
Department of Applied Mathematics and Theoretical
Physics \\
Cambridge University \\
Silver Street \\
Cambridge, CB3 9EW \\
U.\ K.
\end{center}

\begin{center}\it$^c$
Joseph Henry Laboratories\\
Princeton University\\
Princeton, NJ 08544
\end{center}

\begin{center}
November 1997
\end{center}

\begin{abstract}
The BPS spectrum of type I' string theory in a generic
background is derived using the duality with the
nine-dimensional heterotic string theory with Wilson lines.
It is shown that the corresponding mass formula has a natural
interpretation in terms of type I', and it is demonstrated that the
relevant states in type I' preserve supersymmetry.
By considering certain BPS states for different  Wilson lines
an independent confirmation of the
string creation phenomenon in the D0-D8 system is found. We also
comment on the non-perturbative realization of gauge enhancement in
type I', and on the predictions for the quantum
mechanics of type I' D0-branes.
\end{abstract}

\vfill
\setcounter{footnote}{0}
\def\thefootnote{\arabic{footnote}}
\newpage

\renewcommand{\theequation}{\thesection.\arabic{equation}}

\section{Introduction}
\setcounter{equation}{0}

An interesting phenomenon involving crossing D-branes has recently
been discovered \cite{bdg,dfk,bgl}. When a D$p$-brane and a
D$(8-p)$-brane that are mutually transverse cross, a fundamental
string is created between them. This effect is related by U-duality
to the creation of a D3-brane when appropriately oriented
Neveu-Schwarz and Dirichlet 5-branes cross \cite{hw}. These, and all
other brane-creation phenomena can be obtained from M-theory, where
crossing M5-branes lead to the creation of an M2-brane
\cite{dfk,dealwis}. A M(atrix) theory origin of this phenomenon has also
been suggested recently \cite{hwotz}.

An example of a system exhibiting the string creation effect is the
D0-D8 system in type IIA string theory. Unlike the other
D$p \perp$D$(8-p)$ systems, this system is further constrained by
charge conservation to include vacuum strings between the branes
\cite{polstrom,bgl}. The resulting force between the D0- and the
D8-brane vanishes, and the system is in a BPS state \cite{dfk,bgl}.
This system is
closely related to the 9-dimensional type I' string theory \cite{dlp}
which is T-dual to type I on $\bbbr^9\times S^1$. As type I string
theory is unoriented, type I' string theory contains two orientifold
fixed planes, and the theory lives on
$\bbbr^9\times S^1/\bbbz_2$, where the interval $S^1/\bbbz_2$ is
bounded by the two orientifold planes. In addition, the $SO(32)$
Chan-Paton factors of type I (32 D9-branes) become sixteen parallel
D8-branes in type I' (together with their sixteen images on the other
side of the fixed planes), and their positions in the interval are
determined by a Wilson line in the type I theory. The conserved
charges of type I string theory are Kaluza-Klein (KK) momentum and
D-string winding number, and those of type I' are winding number and
D-particle number. The system contains D0- and D8-branes as well as
fundamental strings, and the above effect should therefore play an
important role in the non-perturbative regime of the theory.

Type I string theory has been conjectured to be S-dual to the
$Spin(32)/\bbbz_2$ heterotic string theory \cite{wit}. The map
relating the two theories compactified on $\bbbr^9\times S^1$
is given by \cite{polwit}
\be
\label{hettypeI}
  \lambda_I = {1\over\lambda_h} \,, \qquad
  G_{MN}^I = {G_{MN}^h\over \lambda_h} \,, \qquad
  R_I =  {R_h\over\sqrt{\lambda_h}}   \,,
\ee
where $G_{MN}$ is the ten-dimensional metric ($M,N=0,\ldots,9$),
$\lambda_h$ and $\lambda_I$ are the heterotic and type I coupling
constants, and the radius of the circle is $R_h$ and $R_I$
when measured in the heterotic and type I metrics, respectively. It is
easy to see from these relations that heterotic KK momentum is mapped
to type I KK momentum, and that heterotic winding
number is mapped to D-string winding number in type I.
On the other hand, the winding number of the fundamental string
in type I is not a conserved quantity, and it does not have a dual
description in the heterotic string theory. Combining the above
S-duality map with the T-duality map relating type I and type I',
\be
  \lambda_{I'} = {\lambda_I\over R_I}  \,, \qquad
  G_{MN}^{I'} = G_{MN}^I \,, \qquad
  R_{I'} = {1\over R_I} \,,
\label{simpleTduality}
\ee
we find that the heterotic and type I' parameters are related as
\be
  \lambda_{I'} = {1\over \sqrt{\lambda_h} R_h} \,, \qquad
  G_{MN}^{I'} = {G_{MN}^h\over\lambda_h} \,, \qquad
  R_{I'} = {\sqrt{\lambda_h}\over R_h} \,.
\label{simpleSduality}
\ee
In particular, the heterotic KK momentum is mapped to
type I' winding, and heterotic winding is mapped to the D0-brane
number in type I'. The KK momentum is not a conserved quantity in type
I' (it is related to the winding of the fundamental string in type I),
and it therefore has no heterotic dual.

The simple relations in (\ref{simpleTduality}) and
(\ref{simpleSduality}) hold only when eight D8-branes
are fixed at the location of each orientifold plane, which corresponds
to the Wilson line $A=(0^8,(1/2)^8)$, breaking $SO(32)$ to
$SO(16)\times SO(16)$. In this case, the type I' background metric is
flat and the dilaton is constant, and it is therefore possible to
define a (global) string coupling constant as
$\lambda_{I'} \equiv \exp{(\phi_{I'})}$. This case is also special in
the sense that the 9-dimensional $Spin(32)/\bbbz_2$ heterotic theory
for which the gauge group has been broken to $SO(16)\times SO(16)$ is
T-dual to a suitable compactification of the  $E_8\times E_8$
heterotic string theory (where again the gauge group has been broken
to $SO(16)\times SO(16)$) \cite{ginsparg}.
This in turn implies a duality between the
$E_8\times E_8$ heterotic theory and type I', which has been explored
in \cite{ks,lowe}. From the point of view of M-theory, this duality is
simply the exchange of two compact directions, one of which is $S^1$
and the other $S^1/\bbbz_2$. This has in turn been exploited to
construct the heterotic matrix theory \cite{heteroticmatrix}.
\smallskip

For Wilson lines other than $(0^8,(1/2)^8)$ the situation
is more complicated. In particular, the dilaton of type I' is then not
constant along the interval, the metric is curved, and the simple
duality relations (\ref{simpleTduality}) and (\ref{simpleSduality})
no longer hold. Furthermore, since the two heterotic theories are
not related by T-duality in the general case, type I' is
then not related to the $E_8\times E_8$ theory, and therefore has no
simple M-theory interpretation.

The Wilson lines $(0^{8-N},(1/2)^{8+N})$
(which in type I' correspond to backgrounds with $(8-N)$
D8-branes at one fixed plane and $(8+N)$ at the other) were partially
explored in \cite{polwit}. The heterotic theories in this class
exhibit, at particular radii (which depend on $N$), gauge enhancement
to exceptional groups, and it was shown that this occurs
whenever the type I' theory develops a region where the coupling
diverges. It is then possible that additional states become massless,
thereby avoiding a potential contradiction of the duality with the fact
that type I' (or type I) does not possess a perturbative gauge
enhancement mechanism.

On the other hand, the precise form of the duality map was not
analyzed in \cite{polwit}, and it is clear that it cannot be as simple
as in the $N=0$ case. In particular, type I' winding is quantized in
multiples of $1/2$, as a single open string which stretches from a
D8-brane at one end to a D8-brane at the other end has winding number
$1/2$, and a single closed string which winds around the whole
interval has winding number $1$. The heterotic KK momentum,
on the other hand, is shifted due to the presence of the
Wilson line, and it is not always half-integral. In fact, for
$N=0\,\mbox{mod}\, 4$ the heterotic KK momentum is half-integral, but
for $N=2\,\mbox{mod}\, 4$, it is of the form $n/4$, where $n\in\Zop$,
and for odd $N$, it is of the form $n/8$. If the simple duality map
were true, this would give unphysical values for the type I' winding
number. It is one of the aims of this paper to derive the correct
duality map in these cases, and to demonstrate that it gives rise to
type I' winding which is always half-integral. The critical
observation is that the heterotic KK momentum is mapped to a state
which has, in addition to the D0-branes, a certain amount of type I'
winding.

We will also show that the actual value of the type I' winding number
is somewhat ambiguous as it depends in general on the positions of
the D0-branes
along the interval. In particular, the difference in mass of a
D0-brane as it changes its position can be absorbed into a shift of
the winding number, and this can be understood as a consequence of the
macroscopic D0-D8 strings, which were shown to be present in
\cite{polstrom,bgl}. Furthermore, the creation of D0-D8 strings can be
demonstrated by analyzing the system as a function of the positions of
the D8-branes; our analysis of the duality map between BPS states in
the heterotic and type I' string theories therefore gives further (and
independent) evidence for this phenomenon.

It will also become apparent that the relevant BPS states in
type I' preserve supersymmetry. In the presence of D0- and
D8-branes, this amounts to the condition that the string has a
definite orientation, and we demonstrate that this follows from
the BPS condition of the heterotic theory \cite{dh}.
This latter condition requires that the
right-moving oscillators are in their ground states, and for non-zero winding
number ({\it i.e.} in the presence of D0-branes) this roughly implies
that KK momentum can only flow in one direction, and therefore that
type I' strings can only wind in one direction.
\medskip

The paper is organized as follows. In section~2 we review
nine-dimensional heterotic string theory and gauge enhancement. In
section~3 we derive the precise heterotic-type I' duality map,
including both heterotic momentum and winding, and show that the type
I' winding is always physical and supersymmetric, {\it i.e.}
half-integer and non-positive. In section~4 we use the duality to
exhibit the creation of D0-D8 strings by following the behavior of
certain states as the Wilson line is varied,
and discuss the role these strings play in gauge enhancement.
In section~5 we present
some conclusions, and comment on the predictions of the present
analysis with regards to the D0-brane quantum mechanics.
We have included two short appendices:
in the first we give a table of the lightest type I' BPS states for a
small number of D0-branes, and in the second we generalize the results
of section~3 to Wilson lines corresponding to bulk D8-branes.

While this paper was being finalized, a paper \cite{mnt} appeared
where some results which overlap with section~3 are obtained.

\section{Review of $D=9$ heterotic string theory}
\setcounter{equation}{0}

Let us consider the $Spin(32)/\bbbz_2$ heterotic string theory
compactified on $S^1$ with radius $R$ and a background gauge field
$A$. The left- and right-moving momenta are
\begin{eqnarray}
p_L &=& \left( \sqrt{2\over \alpha_h^\prime}(P + w_hA),
       {n_h  - A\cdot P - w_hA^2/2\over R_h} +
       {w_hR_h\over\alpha_h^\prime}\right)\nonumber \\
p_R &=& {n_h  - A\cdot P - w_hA^2/2\over R_h} -
                 {w_hR_h\over\alpha_h^\prime} \,,
\label{momenta}
\end{eqnarray}
where $P$ is an element of the $Spin(32)/\bbbz_2$ lattice $\Gamma^{16}$,
and $n_h$ and $w_h$ are the KK momentum and winding numbers,
respectively. The mass spectrum is given by
\be
M_h^2 = {1\over 2}p_L^2  + {2\over\alpha_h^\prime}(N_L -1) +
       {1\over 2}p_R^2  +{2\over\alpha_h^\prime}(N_R -c)\,,
\ee
where $N_R,N_L$ are the right and left-moving oscillator numbers,
and $c=0$ or $1/2$ depending on whether the right-moving fermions are
periodic (R) or anti-periodic (NS). Physical states satisfy the level
matching condition
\be
p_L^2  + {4\over\alpha_h^\prime}(N_L -1)  =
            p_R^2  +{4\over\alpha_h^\prime}(N_R -c) \,,
\ee
and BPS states are given by the further requirement that $N_R=c$. The
condition for states to be physical and BPS-saturated is thus
\be
 p_L^2 - p_R^2 = {4\over \alpha_h^\prime} (1 - N_L) \,,
\ee
which can be simplified to
\be
  {1\over 2}P^2 + n_hw_h = 1 - N_L  \,.
\label{physicalBPS}
\ee
In this case, the mass formula becomes
\be
M^2_{h,BPS} = p_R^2 =
    \left( {n_h - A\cdot P - w_hA^2/2\over R_h}
          - {w_hR_h\over \alpha_h^\prime} \right)^2 \,.
\label{hetBPS}
\ee
For example, states with  $N_L=1$ and $P^2=0$ include the gravity
multiplet and the vector multiplets associated with the Cartan
generators of the gauge group. Additional massless vectors, associated
to the roots of the underlying gauge group, have $N_L = 0$, and
therefore satisfy
\be
 p_R^2 = 0 \,, \qquad
 p_L^2 = {4 \over \alpha_h^\prime} \,.
\ee
At zero winding number this gives the roots of the subgroup of $SO(32)$
which is left unbroken by the Wilson line, and for a generic radius
$R_h$ there are no further massless gauge bosons. However, if the radius
satisfies
\be
 R_h^2 = \alpha_h^\prime(1 - A^2/2 ) \; ,
\label{radius}
\ee
where $A^2$ is the length of the shortest vector of the form
$A+\lambda$ where $\lambda\in\Gamma^{16}$, there exist additional
massless vectors for non-zero winding number which give rise to an
`enhancement' of the gauge symmetry.

In this paper we shall be mostly interested in the case where the
Wilson line is of the form $A_{N} = (0^{8-N},(1/2)^{8+N})$ where
$N=0,\ldots,8$. This class of theories exhibits an interesting pattern
of gauge enhancements (table 1).\footnote{There are other enhanced
groups as well, such as $SO(34)$ when $A=(1,0^{15})$ (free-fermionic
point), and SU(18) when $A=(r^{16})$ with $0<r<1/2$.} At a generic
radius, the unbroken gauge group is
$SO(16-2N)\times SO(16+2N)\times U(1)^2$. For $N>0$, the smaller $SO$
group, together with one of the $U(1)$'s is enhanced at the critical
radius
\be
R_h^c(N) = \sqrt{\alpha'_h N \over 8}
\label{critical}
\ee
to $E_{9-N}$. On the other hand, for $N=0$, the critical radius is
$R_h^c(0)=0$, and both  $SO(16)$ groups are enhanced to $E_8$.
In this case all winding numbers contribute to the gauge enhancement.
This reflects the fact that a new continuum degree of freedom has
appeared; indeed, by T-duality, the $N=0$ theory at $R_h=0$ is really
the ten-dimensional $E_8\times E_8$ heterotic string theory
\cite{ginsparg}.

\begin{table}[hbt]
\begin{center}
\begin{tabular}{|c|c|c|l|c|c|c|} \hline
 $N $ & $G$ & $R_h^2/\alpha_h^\prime$ & $G_{enhanced}$
     & $w_h$ & $p_h$ &
        $G$ charges \\ \hline
 $0$ & $SO(16)\times SO(16)$ & $0$ &
       $E_8\times E_8$ & $\bbbz$ & $0$ &({\bf 128},{\bf 1}),
     ({\bf 1},{\bf 128}) \\
 $1$ & $SO(14)\times U(1)$ & $1/8$ &
     $E_8$ & $\pm 1, \pm 2$ & $\pm 1/8, \pm 1/4$ &
      ${\bf 64}_+,{\bf 64'}_-,{\bf  14}_+,{\bf 14}_-$ \\
 $2$ & $SO(12)\times U(1)$ & $1/4$ &
     $E_7$ & $\pm 1, \pm 2$ & $\pm 1/4, \pm 1/2$ &
     ${\bf 32}_+,{\bf 32}_-, {\bf  1}_+,{\bf 1}_-$ \\
 $3$ & $SO(10)\times U(1)$ & $3/8$ &
     $E_6$ & $\pm 1$ & $\pm 3/8$ &
    ${\bf 16}_+,{\bf 16'}_-$ \\
 $4$ & $SO(8)\times U(1)$ & $1/2$ &
     $E_5$ & $\pm 1$ & $\pm 1/2$ &
    ${\bf 8}_+,{\bf 8}_-$ \\
 $5$ & $SO(6)\times U(1)$ & $5/8$ &
     $E_4$ & $\pm 1$ & $\pm 5/8$ &
    ${\bf 4}_+,{\bf 4'}_-$ \\
 $6$ & $SO(4)\times U(1)$ & $3/4$ &
     $E_3$ & $\pm 1$ & $\pm 3/4$ &
     ${\bf 2}_+,{\bf 2}_-$ \\
 $7$ & $SO(2)\times U(1)$ & $7/8$ &
     $E_2$ & $\pm 1$ & $\pm 7/8$ &
    ${\bf 1}_+,{\bf 1}_-$ \\
 $8$ & $U(1)$ & $1$ &
     $E_1$ & $\pm 1$ & $\pm 1$ &
      $+,-$ \\ \hline
\end{tabular}
\caption{Gauge enhancement in $D=9$ heterotic string theory.
Only the relevant part of the gauge group is presented,
and $p_h=n_h-A\cdot P-w_hA^2/2$.}
\end{center}
\end{table}

The cases $N=1,2$ are different from $3\leq N \leq 8$, in that the
$w_h=\pm 2$ states contribute to the gauge enhancement as well as
the $w_h=\pm 1$ states. Indeed, for $N\geq 3$ the enhancement from
$SO(16-2N)\times U(1)$ to $E_{9-N}$ only requires the spinor
representation to become massless, whereas for $N=1,2$
the vector and singlet representation, respectively, has
to become massless as well.

\section{Duality map and type I' BPS states}
\setcounter{equation}{0}

Let us now analyze the duality between the above heterotic theories
and the type I' theory where $(8-N)$ D8-branes are at $x^9=0$ and
$(8+N)$ D8-branes are at $x^9=2\pi$. First, we need to fix some
conventions.

We assume (as in \cite{polwit}) that the D8-brane R-R charge is
$-8\mu_8$ just to the right of the orientifold plane at $x^9=0$,
and that it jumps by $+\mu_8$ each time we cross a D8-brane
to the right. The charge is then $-N\mu_8$ in the bulk of the interval.
\smallskip

The D0-D8 system preserves $1/4$ of the supersymmetry. Let us
denote by $Q_L$ and $Q_R$ the two sixteen-component supercharges
of type IIA string theory. A D8-brane that extends along the
$x^0, \ldots, x^8$ directions (and is a point in
$x^9$) is invariant under the linear combination
$\epsilon_L Q_L + \epsilon_R Q_R$, where
\be
  \epsilon_L = \Gamma_0 \cdots \Gamma_8 \epsilon_R \,,
\label{D8}
\ee
and a D0-brane is invariant under the linear combination with
\be
 \epsilon_L = \Gamma_0 \epsilon_R \,.
\label{D0}
\ee
Using $\Gamma_0^2=1$, the two equations (\ref{D8}) and (\ref{D0}) are
solved by $\epsilon_L=\Gamma_0\epsilon_R$ provided $\epsilon_R$
satisfies
\be
\epsilon_R = \Gamma_1 \cdots \Gamma_8 \epsilon_R \,,
\label{D08}
\ee
and it is easy to check that such $\epsilon_R$ exist.

On the other hand, an elementary string that is stretched along the
$x^9$ axis is invariant under the supersymmetry generator provided
that
\be
 \epsilon_L = \Gamma_0 \Gamma_9 \epsilon_L \,, \qquad
 \epsilon_R = \pm \Gamma_0 \Gamma_9 \epsilon_R \,,
\label{NS1}
\ee
where the sign depends on the orientation of the string. It is then easy
to check that the three conditions (\ref{D8}), (\ref{D0}) and
(\ref{NS1}) are compatible only for the upper sign in (\ref{NS1});
this means that in a D0-D8 system, supersymmetry is only
unbroken if any strings that are present are oriented in a specific
direction. For definiteness we shall assume that configurations with
unbroken supersymmetry correspond to strings stretching from right
to left for which the winding number is negative.

It should be stressed that this constraint is a consequence of
{\it both} the D0- and the D8-brane being present. In particular, for
states without any D0-branes, there is no restriction on the direction
of the string.
\smallskip

The winding number of type I' is quantized in units of $1/2$, which
corresponds to a string that stretches once across the interval. The
D0-brane number is also quantized in units of $1/2$; the minimal value
corresponds to a single D0-brane that is stuck at one of the
orientifold planes.

\subsection{Type I' BPS spectrum}

Our calculation follows the analysis of \cite{polwit},
but we
find it useful to re-introduce all the dimensionful
parameters
(other than $\hbar$ and $c$ of course). The low energy
effective
actions of the two theories are given (up to terms that
we shall not
need) by
\be
 S_h = \int d^{10}x \sqrt{-G_h} e^{-2\phi_h} \left\{
            {1\over 2\kappa^2} \Big[ \R_h + \cdots \Big]
            - {1\over 8g_{YM}^2} \Tr_V F^2  \right\} \,,
\ee
where the Yang-Mills and gravitational couplings are
related
by
$2\kappa^2 = \alpha_h^\prime g_{YM}^2$ \cite{het}, and
\begin{eqnarray}
 \lefteqn{S_{I'} = {1\over 2\kappa^2} \int d^{10}x
   \sqrt{-G_{I'}}
      e^{-2\phi_{I'}}
            \Big[ \R_{I'} + \cdots \Big]} \nonumber \\
       & &    - T_8 \sum_i \int_{x^9 = x^9_i} d^9 x
           e^{-\phi_{I'}}
          \Tr_f \left[ \sqrt{\det (g^{I'}_{\mu\nu} -
          2\pi\alpha_{I'}^\prime
              F_{\mu\nu})} - \sqrt{-g_{I'}} \right] \,.
\end{eqnarray}
Here the second term is the sum of the DBI actions on the
world-volumes of the D8-branes, $g^{I'}_{\mu\nu}$ is the
nine-dimensional metric, and the D8-brane tension is
given by
$T_8 = \mu_8/(\sqrt{2}\kappa)$ \cite{polchinski_tasi}.
The constants $\kappa$ and $\mu_8$ are fixed by the Dirac
quantization
condition for D-branes, and by the normalization of the
D-string
tension \cite{dealwis2}
\be
 2\kappa^2 = (2\pi)^7(\alpha_{I'}^\prime)^4 \,, \qquad
   \mu_8^2 = (2\pi)^{-9} (\alpha_{I'}^\prime)^{-5} \,.
\label{constants}
\ee
Expanding the DBI action to quadratic order in
$\alpha_{I'}^\prime$ gives
\be
 S_{I'} = {1\over 2\kappa^2} \int d^{10}x \sqrt{-G_{I'}}
      e^{-2\phi_{I'}}
            \Big[ \R_{I'} + \cdots \Big]
    - {(\pi\alpha_{I'}^\prime)^2 \mu_8\over \sqrt{2}\kappa}
      \sum_i \int_{x^9 = x^9_i} d^9 x \sqrt{-g_{I'}}
      e^{-\phi_{I'}}
       \Tr_f F^2   \,.
\ee
For $\kappa =1$ this reproduces precisely the actions
considered in
\cite{polwit}. The type I' background dilaton and metric
are given as
\cite{polwit}
\begin{eqnarray}
 e^{\phi_{I'}(x^9)} & = & \Big[\kappa\Omega(x^9)/C\Big]^5
     \nonumber \\
 G^{I'}_{MN}(x^9) & = &  \Omega^2(x^9) \eta_{MN} \,,
\label{background}
\end{eqnarray}
where
\be
  \Omega(x^9) = \kappa^{-1} C
     [ 3\mu_8 C (B + Nx^9) /\sqrt{2}]^{-1/6}  \,,
\label{Omega}
\ee
and $B$ and $C$ are constants ($B\geq 0$). The relation
between these parameters and the heterotic parameters $R_h$ and
$\phi_h$ is deduced by replacing the non-compact components of
$\eta_{MN}$ ($M,N=0,\ldots 9$) above with a slowly varying function
of the non-compact dimensions $\gamma_{\mu\nu}$
($\mu,\nu=0,\ldots 8$), and comparing the dimensional reductions of
the low energy effective actions. The duality relation involves a Weyl
rescaling of the nine-dimensional heterotic metric,
$\gamma_{\mu\nu} = D^2 g^h_{\mu\nu}$, where $D$ inevitably depends on
$B$ and $C$.
The comparison of the two gravitational actions then leads to
\be
 2\pi R_he^{-2\phi_h} = \kappa^{-10}D^7C^{10}
    \int_0^{2\pi}dx^9 \Omega^{-2}(x^9) \,,
\label{gravity}
\ee
and comparing the Yang-Mills actions gives
\be
 2\pi R_he^{-2\phi_h} = 2^{5/2} \pi^2 \mu_8
    \kappa^{-4} \alpha^{\prime -1}_h
    \alpha^{\prime 2}_I D^5 C^5 \,.
\label{YM}
\ee
Combining (\ref{gravity}) and (\ref{YM}) we then find
\be
 DC^{5/3} = 3^{-2/3} 2^{7/3} \pi \mu_8^{1/3} \kappa^2
\alpha_h^{\prime -1/2}
    \alpha_{I'}^\prime N^{1/2} \Big[(B+2\pi N)^{4/3} -
B^{4/3}\Big]^{-1/2}  \,.
\label{D}
\ee
The duality maps pure momentum states of the heterotic
string ($w_h=0$)
to pure winding states of type I', and so
\be
  1/R_h = D M^{I'}_{winding} \,,
\label{KK}
\ee
where $M^{I'}_{winding}$ is the mass of a type I' string
that stretches
{\it twice} along the interval as measured in the metric
$\gamma_{\mu\nu}$,
\be
 M^{I'}_{winding} =
       {1\over \pi\alpha_{I'}^\prime}
           \int_0^{2\pi} dx^9 \Omega^2(x^9)
       = {1\over D}\sqrt{8\over \alpha_h^\prime N}
          \left[ { (B+2\pi N)^{2/3} - B^{2/3})\over
           (B+2\pi N)^{2/3} + B^{2/3}) } \right]^{1/2} \,.
\label{windingmass}
\ee
Here we have used (\ref{D}) to eliminate $C$. The mass of a
D0-brane which is located at $x^9$, as measured in the same
metric, is given by
\be
  M^{I'}_{D0} (x^9) = {1 \over \sqrt{\alpha_{I'}^\prime}}
            \Omega(x^9) e^{-\phi_{I'}(x^9)}
         = {1\over D} \sqrt{2N\over\alpha_h^\prime}
     { (B + Nx^9)^{2/3}\over \Big[(B+2\pi N)^{4/3} -
    B^{4/3}\Big]^{1/2}}
   \,,
\label{D0mass}
\ee
where we have used both (\ref{constants}) and (\ref{D}).
It is then easy to check that
\be
 M^{I'}_{D0}(2\pi) - M^{I'}_{D0}(0) =
   {N\over 2} M^{I'}_{winding} \,,
\label{D0diff}
\ee
or more generally that the masses of D0-branes which are
located at different values of $x^9$ differ by the
mass of $N$ strings that are stretched between the two points in
$x^9$,
\be
   M^{I'}_{D0}(x^9_1) - M^{I'}_{D0}(x^9_2) =
   N M^{I'}_{string}(x^9_1,x^9_2) \,.
\label{D0diff2}
\ee
This is consistent with the fact that D0-branes feel a linear
gravity-dilaton potential in the presence of D8-branes
\footnote{Actually, since the metric $\gamma$ is flat, the
potential is due only to the dilaton in this metric. In curved
metrics there would also be a gravitational contribution.}.

In remains to analyze the heterotic winding states
whose KK momentum vanishes.
{}From the situation for $N=0$ one may expect
that these states correspond in type I' to states that consist
of D0-branes, but do not have any winding.
However, inverting (\ref{KK}) and using
(\ref{windingmass}) and (\ref{D0mass}),
we find
\be
    R_h/\alpha_h^\prime = D \left[ {1\over 2} M^{I'}_{D0}(0)
      +{N\over 8} M^{I'}_{winding} \right] \,.
\label{winding2}
\ee
This suggests that the heterotic winding states map
to states that consist of D0-branes at $x^9=0$, together with
some winding
\footnote{This result could have been anticipated as follows.
There exist heterotic states with winding number $1$
which become massless at the critical radius. These
must correspond to a type I' configuration of half a D0-brane
at $x^9=0$ without any winding.
Using (\ref{KK}), the heterotic mass formula for this state
at a generic radius then implies (\ref{winding2}).}.
The factor of $1/2$ corresponds to the fact that a single
D0-brane is stuck at the orientifold plane,
and therefore contributes only half of its mass to the bulk.

As it stands, the winding contribution in (\ref{winding2})
is in $\bbbz/8$, but the winding number in type I' must be
quantized in units of $1/2$. If we combine the heterotic momentum
states (\ref{KK}) and the heterotic winding states
(\ref{winding2}) into the BPS mass formula (\ref{hetBPS}), we find
\be
 M^{I'}_{BPS} = \left| w_{I'} M^{I'}_{winding} -
        n_{I'} M^{I'}_{D0}(0) \right| \,,
\label{IABPS1}
\ee
where
\be
 w_{I'} = n_h - A_N\cdot P - w_h(1+N/4) \,, \qquad
 n_{I'} = w_h/2 \,,
\label{IAcharges}
\ee
are the winding number and D0-brane number, respectively. Here we have
assumed that all D0-branes are located at $x^9=0$.

It is easy to see that $w_{I'}$ is now in $\bbbz/2$, unless $N$ and
$w_h$ are odd (in which case it can be of the form $1/4+\bbbz/2$). On the
other hand, because of (\ref{D0diff2}), the winding number is not
unambiguous, and it depends in fact on the precise locations of the
D0-branes. In order to obtain a physical value for the winding number,
we therefore have to assume that, when $w_{I'}$ is of the form
$1/4+\bbbz/2$, half a D0-brane is at $x^9=2\pi$ rather than at
$x^9=0$; this leads to an additional contribution to the winding
number of $N/4$, and the resulting value for $w_{I'}$ is then in
$\bbbz/2$.
\smallskip

More generally, it is clear that the analysis of the mass formula
alone does not predict the actual locations of the D0-branes, and
therefore the actual value of the winding number. In particular, from
the point of view of the mass formula, it is always possible to `move'
whole D0-branes from $x^9=0$ to $x^9=2\pi$ without modifying the
condition $w_{I'}\in\bbbz/2$. There exist, however,
additional constraints that restrict these possibilities. In
particular, as will be discussed in the following subsection, the
condition of unbroken supersymmetry gives a constraint on the winding
number (and therefore on the possible locations of the D0-branes). 
Furthermore, 
for a certain class of heterotic BPS states, the nature
of the gauge charges fixes the locations of the D0-branes; this will
be discussed in section~4.

\subsection{A condition on the winding number}

As we have explained before, the configuration of a D0-brane, a
D8-brane and a fundamental string is only supersymmetric if the string
is oriented in a particular way. We have chosen the convention that
(in the presence of a D0-brane) the winding number must be
non-positive.

It is clear that the BPS states of type I' must satisfy
this requirement. This is not apparent from (\ref{IAcharges}), since
it seems one could get positive winding by choosing $n_{h}$
arbitrarily large, and negative winding by choosing $w_{h}$
arbitrarily large.
However, the values of $n_h$ and $w_h$ are not arbitrary since they
must satisfy the heterotic physical-BPS condition
(\ref{physicalBPS}). We can express this condition in terms of the
type I' variables, and we find that
\be
  2n_{I'}w_{I'} = 1 - {N\over 2} n_{I'}^2
    - {1\over 2}(P + 2n_{I'} A_N)^2 - N_L \,.
\label{condition}
\ee
For $n_{I'}>0$, it follows from (\ref{condition}) that
\be
w_{I'} \leq {1 \over 2 n_{I'}} - {N n_{I'} \over 4 } \,,
\ee
as $N_L\geq 0$. Since $w_{I'}$ is in $\Zop/4$, the only positive values
for $w_{I'}$ can arise if
\be
{1 \over 2 n_{I'}} - {N n_{I'} \over 4 } \geq 1/4 \,.
\ee
This can only happen for $n_{I'}=1/2$ if $N=0,\ldots, 6$, $n_{I'}=1$
if $N=0,1$, and $n_{I'}=3/2,2$ if $N=0$. These cases can be checked
directly, and it turns out that in each case (\ref{condition}) implies
that  $w_{I'}\leq 0$. For the benefit of the reader we have worked out
the maximal value of $w_{I'}$ for a certain class of BPS states in
appendix~A.

We have thus shown that the winding number is always non-positive,
provided that all D0-branes are located at $x^9=0$. It is also easy to
check that for those states where the condition $w_{I'}\in\bbbz/2$
requires that at least half a D0-brane is at $x^9=2\pi$, the modified
winding number (which takes into account that half a D0-brane has been
moved) also satisfies this condition. We can then turn the argument
around, and regard the condition $w_{I'}\leq 0$ as a constraint on the
possible locations of the D0-branes.
\smallskip

Using the fact that $w_{I'}\leq 0$ for BPS states, the BPS mass
formula for type I' then becomes
\be
  M^{I'}_{BPS} = |w_{I'}| M_{winding} + |n_{I'}| M_{D0}(0) 
\label{IABPS2}
\ee
if $w_{I'}\in\bbbz /2$, and 
\be
 M^{I'}_{BPS} = (|w_{I'}| - N/4) M_{winding} + (|n_{I'}|-1/2) M_{D0}(0) 
           + 1/2 M_{D0}(2\pi)
\label{IABPS2p}
\ee
if $w_{I'}\in\bbbz/2 + 1/4$. 
This means that the mass of a type I' BPS state is simply the sum of
the mass associated to the strings and the mass associated to the
D0-branes, as may have been expected. It should be stressed that this
is in marked contrast to the heterotic BPS mass formula (\ref{hetBPS}).

\section{String creation and gauge enhancement}
\setcounter{equation}{0}

In this section we examine how some of the type I' BPS states evolve
as the D8-branes are moved from one end of the interval to the other,
{\it i.e.} as the Wilson line of the heterotic theory is changed. As
we shall see, the winding number of the states in type I' changes as
the D8-branes are moved, and we shall demonstrate that this can be
accounted for in terms of D8-D8 strings that become longer or shorter,
and fundamental strings that are created as D8-branes cross
D0-branes.

\subsection{D0-D8 strings}

Let us begin with the $N=0$ case, for which eight D8-branes
are located at each orientifold plane. Let us further consider
the BPS states that have zero winding number in type I'.
These states were discussed in \cite{ks} from a slightly different
point of view. There are three kinds of states: those charged under
the first $SO(16)$ (``left-charged"), those charged under the second
$SO(16)$ (``right-charged"), and neutral states. The charged states
can be further subdivided according to whether the heterotic winding
and momentum numbers are even or odd (table~2).
\begin{table}[hbt]
\begin{center}
\begin{tabular}{|c|c|c|c|c|} \hline
    $N_L$ & $w_h$ & $n_h$ & $P$ & $P+w_hA$ \\ \hline
    &&&& \\[-12pt]
    $0$ & $2k$ & $-2k$ & $k(0^{8};(-1)^{8})
    \pm (1, \pm1, 0^{6}; 0^{8})$
                & $(\pm1,\pm1,0^6;0^8)$ \\[4pt]
                & $2k+1$ & $-2k-1$
                & $k(0^{8}; (-1)^{8}) + (-\half)^{16}) +
                   (1^{2l},0^{8-2l};0^8)$ 
                & $((\half)^{2l},(-\half)^{8-2l};0^8)$ \\[3pt]
    \hline
    &&&& \\[-12pt]
    $0$ & $2k$ & $-2k\pm 1$
           & $ k(0^{8};(-1)^{8}) \pm 
             (0^{8};1^2,0^6)$
           & $\pm (0^8;1^2,0^6)$ \\[4pt] 
           & $2k$ & $-2k$
           & $k(0^{8};(-1)^{8}) \pm 
             (0^{8};1,-1,0^{6})$
           & $\pm (0^8;1,-1,0^6)$ \\[4pt] 
           & $2k+1$ & $-2k+1-l$
           & $k(0^{8}, (-1)^{8}) - (0^{8}; 1^{2l}, 0^{8-2l})$
           & $(0^8;(-\half)^{2l},(\half)^{8-2l})$ \\[3pt]
    \hline
     &&&& \\[-12pt]
    $1$ & $2k$ & $-2k$ & $-k(0^{8},1^{8})$ &
    $(0^{16})$ \\[4pt] \hline 
\end{tabular}
\caption{$N=0$ heterotic BPS states with $w_{I'}=0$.
The first two rows are left-charged states, the next
three rows are right-charged states, and the last row
corresponds to neutral states. The various $1$'s and $-1$'s
can be located anywhere in the corresponding eight entries,
and $l$ in the second row is a positive integer $\leq 4$.}
\end{center}
\end{table}

The number of D0-branes in these states is equal to $w_h/2$, and the
winding number is zero. We shall only consider those states for which
$w_h>0$; analogous statements hold for the BPS states with $w_h<0$,
which correspond to anti-D0-branes.
The duality with the heterotic theory suggests that the D0-branes form
bound states with the orientifold planes and D8-branes at $x^9=0$ and
$x^9=2\pi$ for the left- and right-charged states, respectively, and
that they form bound states in the bulk for the neutral states.

There exist two mechanisms by means of which gauge charges can arise
in type I': the state can contain open strings that begin or end
on D8-branes, and it may contain one or more D0-branes
that are stuck at an orientifold plane. The former states account for
weights in the conjugacy classes of the identity and vector
representations. For example, open strings that begin and end on
D8-branes which are located at the same orientifold plane correspond
to the roots of the underlying gauge group, and D8-D8 strings which
stretch across the interval correspond to states in the bivector
representation. On the other hand, D0-D8 strings give rise to states
in the vector representation.

A half D0-brane that is stuck on an orientifold plane generates,
through its fermionic zero modes, states that transform under the
corresponding spinor representation. If there are more such
D0-branes on one of the orientifold planes, this gives rise to states
that transform in the symmetric product of the spinor representation.

We shall assume that in any of the above BPS states {\em at most}
half a D0-brane is stuck at an orientifold plane. In particular, this
implies that weights in the conjugacy classes of the identity and the
vector representations arise solely from open strings, whereas weights
in the conjugacy classes of the spinors arise from a combination of
open strings and half a D0-brane.

Consider for example the states with weight vectors
$\pm(1,\pm 1,0^6;0^8)$. In the type I' picture these charges arise from
strings between the corresponding D8-branes. The overall sign
corresponds to the orientation of the string, and the
relative sign is $-$ for a string between two D8-branes in the
interval $0\leq x^9\leq 2\pi$, and $+$ for a string between a
D8-brane and the {\em image} of another. From the point of view of the
interval $0\leq x^9\leq 2\pi$, the string in the second case begins on
a D8-brane, is reflected by the orientifold, and ends on another
D8-brane (figure~1).
It should be mentioned, that even in the presence of D0-branes, both
orientations of the string preserve supersymmetry as the string has
vanishing length.
\begin{figure}[htb]
\epsfysize=4cm
\centerline{\epsffile{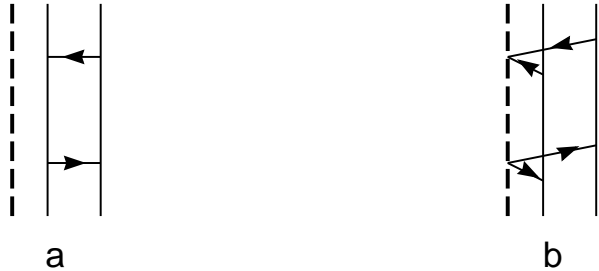}}
\caption{States corresponding to the weights (a) $\pm(1,-1)$, and
(b) $\pm(1,1)$. The dashed vertical line represents an orientifold
8-plane and the solid vertical lines represent D8-branes.}
\end{figure}

We want to vary the Wilson line in such a way as to move
one of the D8-branes from $x^9=0$ to $x^9(r)$, so
\be
  A = (0^8; (1/2)^8) + \Delta A\,, \qquad
\Delta A = (0^7, r ; 0^8) \,,
\label{changewilson}
\ee
where $r$ can be in any of the first eight entries.
The precise $r$ dependence of $x^9(r)$ is rather involved,
and the duality map in this case is worked out in appendix B.
When the D0-branes are near $x^9=0$, the map is
given by
\be
w_{I'}=n_{h}-A\cdot P-w_h(1 + r/2)\,, \qquad \qquad
n_{I'} = w_h/2 \,,
\label{w0}
\ee
and when the D0-branes are near $x^9=2\pi$ the map changes to
\be
w_{I'}=n_{h}-A\cdot P-w_h \,, \qquad \qquad n_{I'} = w_h/2 \,.
\label{w2pi}
\ee
Let us describe the change in $w_{I'}$ for the different types of states
in turn.

\begin{description}
\item[Left-charged states, $w_h$ even] \mbox{}\\
   For a given value of $k$ and two directions in the root lattice of
   the (left) $SO(16)$, there are four states that correspond to the
   four different choices of sign. In type I', the two directions
   correspond to the choice of two D8-branes, and the four states
   correspond to the various open strings between them.
   If $\Delta A$ is in one of the other six
   directions, $A\cdot P$ does not change
   as $A\mapsto A+\Delta A$; in type I' this corresponds
   to the situation where one of the other D8-branes is moved.
   If $\Delta A$
   is in the direction of one of the two given lattice directions then
   $-A\cdot P$ changes by $\pm r$ . In type I'
   this corresponds to moving one of the two chosen D8-branes, which
   increases
   the length of the string and therefore contributes $\pm r$ to the
   winding number. Here the sign depends on the orientation, and there
   are therefore two states for which the sign is plus, and two
   for which it is minus.

   This appears to be in contradiction with the requirement
   that the string have a negative orientation to preserve
   supersymmetry in the presence of D0-branes.
   In the initial $N=0$ background the contradiction is
   avoided since the strings have vanishing length,
   and in fact the two orientations are related by a gauge
   transformation. On the other hand, as the string becomes longer,
   the states with positive orientation would appear to break
   supersymmetry.

   However, there is an additional contribution to the change in
   $w_{I'}$ given by $-w_hr/2$. This contribution is independent of
   the direction of $\Delta A$. In the type I' picture it corresponds to
   $w_h/2$ D0-D8 strings, which are created when the D8-brane
   crosses the D0-branes. The orientation of these strings is fixed
   by the sign of the background R-R 8-brane charge
   \cite{bgl}, and it is negative in our conventions.

  As a consequence, if $\Delta A$ corresponds to moving one of the two
  selected D8-branes, both effects will contribute. In the two states
  for which the D8-D8 string is positively oriented, this
  string will combine with a negatively oriented D0-D8 string to give
  a positively oriented D0-D8 string of vanishing length (figure~2(a-c)),
  thereby avoiding the possible contradiction with supersymmetry,
  which is manifest in the heterotic picture.\footnote{In the final
  configuration there is a non-vanishing force on the D0-brane
  towards $x^9=0$, and therefore classically the length of the string
  is indeed zero. In quantum mechanics the picture is less clear.}
  This cancellation has important consequences for gauge enhancement,
  as we shall soon see. For the other two states the orientations are
  both negative, and the winding numbers add to a net negative
  number (figure~2(d)).
\begin{figure}[htb]
\epsfysize=7.9cm
\centerline{\epsffile{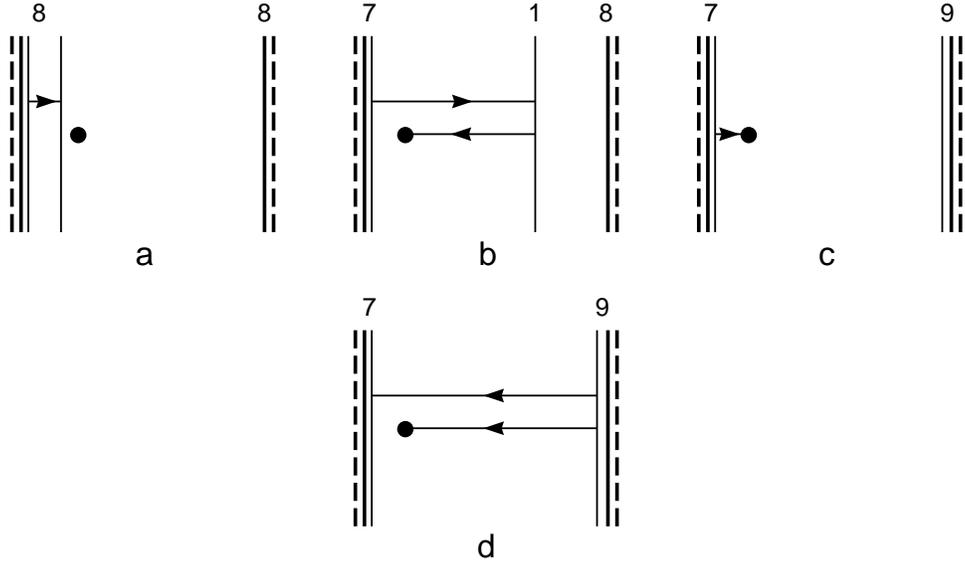}}
\caption{Moving a D8-brane for left-charged states.
(a) represents a state with weight $(+1,-1,0^6;0^8)$ and one
    D0-brane in the background of eight D8-branes on each side.
(b) As the D8-brane associated with the $-1$ entry moves to the
    right, the D8-D8 string becomes longer, and a D0-D8 string of
    opposite orientation is created.
(c) The two strings combine into a D0-D8 string of vanishing length,
    resulting in a net constant force towards $x^9=0$.
(d) As the D8-brane associated with the $+1$ entry moves to the
    right, the D8-D8 string becomes longer, and a D0-D8 string of the
    same orientation is created; there is no force in this case.}
\end{figure}

\item[Left charged states, $w_h$ odd] \mbox{}\\
  For these states, half a D0-brane is stuck at the orientifold plane at
  $x^9=0$ and contributes a weight in the spinor representation. In
  addition there are $2l$ D8-branes at $x^9=0$ that support D8-D8
  strings. The correspondence between the two terms in (\ref{w0}) and
  the number of D8-D8 and D0-D8 strings is not as simple as in the
  previous case. In fact $-A\cdot P$ changes by $\mp  r/2$, depending
  on whether $\Delta A$ is in one of the $2l$ special directions or
  not. By itself, this change cannot correspond to a growing D8-D8
  string, which would contribute $\mp r$ to the winding. However, the
  additional change in $w_{I'}$ from the term $-w_hr/2$ can be broken
  into two parts
  \be
    -w_hr/2 = -kr - r/2 \,.
  \ee
  The first part corresponds to the D0-D8 strings that are created
  when the D8-brane crosses the $k$ D0-branes which live just off
  the orientifold plane. The second part does not correspond to a
  D0-D8 string, since the D8-brane never crosses the additional $1/2$
  D0-brane\footnote{In fact, there are no D0-D8 strings associated
  with this $1/2$ D0-brane, as it is confined to precisely $x^9=0$,
  where the value of the R-R 8-brane charge is $0$.},
  but rather combines with the $-A\cdot P$ contribution to
  give a D8-D8 string for the case when one of the $2l$ special
  D8-branes is moved (figure~3), and to cancel the above winding when
  one of the other D8-branes is moved.
\begin{figure}[htb]
\epsfysize=4cm
\centerline{\epsffile{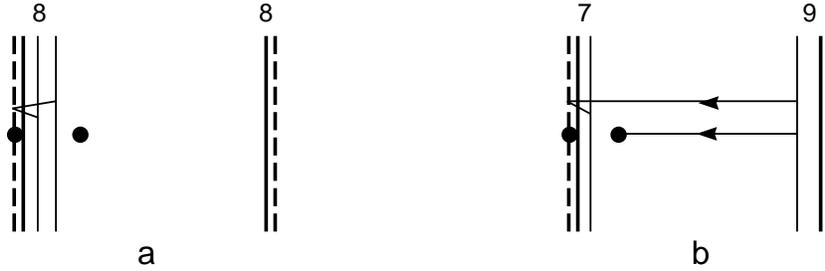}}
\caption{Moving a D8-brane in left-charged states.
(a) represents a state with weight $((1/2)^2,(-1/2)^6;0^8)$ 
    ($l=1$) 
    with $3/2$ D0-branes.
(b) As one of the two special D8-branes is moved to the right, 
a D8-D8 string of 
negative orientation becomes longer, and adds to the D0-D8 string
of negative orientation.}
\end{figure}

\item[Right-charged states] \mbox{}\\
  In this case the type I' winding number is given by
  (\ref{w2pi}), where only the term $-A\cdot P$ can change.
  In fact this term does not change, since the first eight entries
  of $P$ are zero. In the type I' picture the winding remains
  unchanged since there are no D8-D8 strings at $x^9=0$,
  and D0-D8 strings are not created until the D8-brane
  reaches $2\pi$, which is where the D0-branes are
  (see figure~4). These strings will therefore have vanishing
  length, and will not contribute to the winding. They will however
  change the weight vector, consistent with the heterotic picture.
\begin{figure}[htb]
\epsfysize=4cm
\centerline{\epsffile{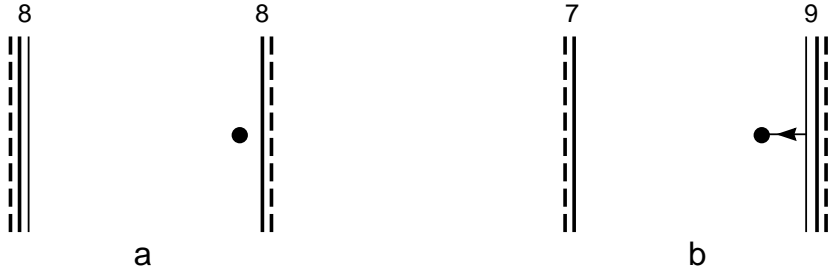}}
\caption{Moving a D8-brane in right-charged states.
(a) represents any right-charged state with one D0-brane. All
    open strings are between D8-branes on the right.
(b) As a D8-brane is moved to the right, there is no effect until it
    crosses the D0-brane at $2\pi$, when a D0-D8 string of vanishing
    length appears. The net force on the D0-brane vanishes in this
    case.}
\end{figure}

\item[Neutral states] \mbox{}\\
  In this case there are no D8-D8 strings in the type I' picture.
  Any change in the type I' winding should therefore arise from
  D0-D8 strings. The winding number is somewhat ambiguous
  in this case, as the D0-branes are not localized at either end.
  We can however define $w_{I'}$ as if the D0-branes were all at
  $x^9=0$. Eq.~(\ref{w0}) then shows that the change in winding
  corresponds precisely to the creation of $w_h/2$ D0-D8 strings.
\end{description}
\goodbreak
To conclude, we see that the creation of D0-D8 strings
is necessary in order to explain the evolution of type I' BPS
states, given the duality with heterotic string theory.
We have assumed that, except for a half D0-brane
in the case when there are $\bbbz +1/2$ D0-branes, D0-branes are
not stuck at the orientifold, but rather are bound (together
with their images) to the orientifold plane and D8-branes, and this
has allowed us to explain the evolution of the above states
satisfactorily.

\subsection{Gauge enhancement}

We are now in a position to understand how the gauge enhancement
pattern of section 2 is realized non-perturbatively in type I'.
For every $N$ there is a point in the moduli space of type I',
corresponding to $B=0$, 
where
the dilaton (\ref{background}) blows
up at $x^9=0$, and D0-branes located at $x^9=0$ become massless
(\ref{D0mass}). This point corresponds to the appropriate critical
radius (\ref{critical}) in the heterotic picture \cite{polwit}. The
case $N=0$ has already been explained in \cite{ks}. In all the other
cases, the states in table 1 with $w_h=1$ (or $-1$) correspond to type
I' states consisting of $1/2$ D0-brane (or $1/2$ anti-D0-brane) at
$x^9=0$, and no winding ({\it e.g.} figure~5(a)).
The mass of these states is thus entirely due
to the $1/2$ D0-brane at $x^9=0$, and it therefore vanishes for $B=0$.
These states transform in the spinor representation of
of the gauge group $SO(16-2N)$, and are of course spacetime
vector multiplets.

For $N=1,2$ there are additional massless vectors with
$w_h=\pm 2$. In type I' these are states that consist of a whole
D0-brane without any net winding, and are therefore massless when
$B=0$. It follows from the analysis of the heterotic theory that
these states are only BPS if $N=1,2$, but not for $N\geq 3$. This can
be understood from the point of view of type I' as follows. Let us
consider the $w_h=2$ state in the $N=0$ background which is
charged under the left
$SO(16)$, and which becomes massless
at the critical (vanishing) radius.
This state transforms in the adjoint of $SO(16)$.
In type I' it consists
of one whole D0-brane near $x^9=0$
without any winding,
and therefore belongs to
the first class of states discussed above.
In fact figure~2(a) depicts one of its components.

As we move one of the two D8-branes to $2\pi$,
leaving seven D8-branes at $0$, the gauge group at
$x^9=0$ becomes $SO(14)$. As discussed above, 
for the two components of the original state for which the D8-D8
string is positively oriented, the net winding number will vanish as a
consequence of the cancellation between the D8-D8 string
and the D0-D8 string. 
These states will therefore become
massless at the new ($N=1$) critical point. Both components
will contain a short D0-D8 string (figure 5(b,c)), and will
therefore give rise to a state in the vector ({\bf 14}) representation
of $SO(14)$. Together with the $1/2$ D0-brane state
which gives rise to the spinor ({\bf 64}) of $SO(14)$, and with
the $1/2$ and $1$ anti-D0-brane states, we get an enhancement of
$SO(14)\times U(1)$ to $E_8$.
\begin{figure}[htb]
\epsfysize=4cm
\centerline{\epsffile{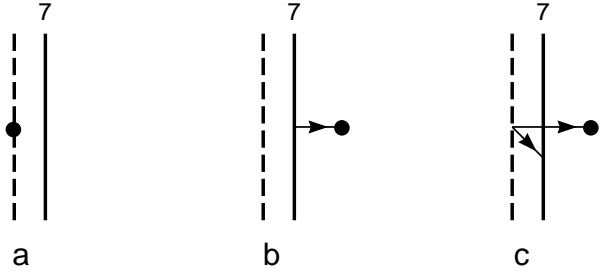}}
\caption{$SO(14)\times U(1)\rightarrow E_8$.
(a) spinor ${\bf 64}_+$
(b) and  (c) give rise to the vector representation ${\bf 14}_+$.}
\end{figure}

Next let us move the other D8-brane, {\it i.e.} the D8-brane
in figure~5(b,c) on which the string ends 
(or begins); the result is shown in figure~6(b,c). The state in
figure~6(b) has winding and is therefore always massive, but the state
in figure~6(c) has no winding, and thus becomes massless at the new
($N=2$) critical point. This state contains two short strings which
are understood to stretch between the D0-brane and its image. (If we
restrict attention to the interval $0\leq x^9\leq 2\pi$ then this is
represented as in figure~6(c)). The state is therefore neutral with
respect to the $SO(12)$ gauge group at $x^9=0$, but it is of course
charged under the R-R $U(1)$. 
Together with the spinor ({\bf 32}) from the $1/2$ D0-brane,
and the anti-D0-brane states, we then get an enhancement
to $E_7$.
\begin{figure}[htb]
\epsfysize=4cm
\centerline{\epsffile{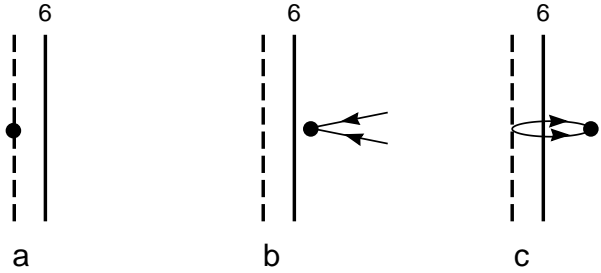}}
\caption{$SO(12)\times U(1)\rightarrow E_7$
(a) spinor ${\bf 32}_+$.
(b) is massive, and
(c) is a singlet ${\bf 1}_+$.}
\end{figure}

If we move any further D8-branes we will not lengthen any
D8-D8 strings, but D0-D8-strings  will still be created. As a
consequence, the above states will acquire non-vanishing
winding, and therefore will never be massless. We can therefore
conclude that states with $n_{I'}=1$ only contribute to
gauge enhancement for $N=1,2$; this is in agreement with
the dual heterotic picture.

\section{Conclusions}

In this paper we have analyzed the duality between the
nine-dimensional heterotic and type I' string theories for a large class
of Wilson lines. We have given the duality map between the relevant
BPS states in detail, and we have shown that the mass formula for type
I' states has a simple interpretation in terms of D0-branes and
winding strings. We have also shown that the heterotic BPS condition
implies that, in the presence of D0-branes, the winding number of type
I' has a definite sign; this is in accord with the fact that in the
presence of D8- and D0-branes, supersymmetry is only unbroken if any
fundamental string has a definite orientation.

A crucial element in our discussion was the realization that a
heterotic winding state corresponds to a certain number of D0-branes
together with some winding. We have also seen that the difference in
mass between D0-branes at different positions along the interval
corresponds to the mass of a string that stretches between these
positions. We have analyzed a class of BPS states as one of
the D8-branes is moved from one side of the interval to the other, and
we have found independent evidence for the proposal that a fundamental
string is created whenever a D8-brane crosses a D0-brane. Our analysis
also suggests that, at least for the states in section~4,
at most half a D0-brane is stuck at the orientifold plane, and that
the remaining D0-branes are bound to the orientifold planes and
D8-branes without being constrained to move on the orientifold plane.

The analysis also predicts that certain bound states should exist
in the quantum mechanics of D0-branes in an $N\neq 0$ type I'
background, and it would be interesting to check this directly. In
particular, bound states of D0-branes without any winding are absent,
with the only exception of two D0-branes for $N=1$ and $N=2$. (In these
cases there exist however short strings which 
provide a constant attractive force).
This is in marked contrast to the $N=0$ background, where bound states
of any number of D0-branes are predicted to exist.
Since these bound states are necessary for an
M-theory interpretation (or matrix theory formulation), we expect
neither to be applicable to the $N\neq 0$ backgrounds. This is also
consistent with the fact that D8-branes, by themselves, have no
M-theory interpretation; only the combination of eight D8-branes and
an orientifold plane is known to arise in M-theory.

\section*{Acknowledgments}

We would like to thank Michael Green, Igor Klebanov and Juan Maldecena
for useful discussions.
O.B. is supported in part by the NSF under grants PHY-93-15811
and PHY-92-18167, M.R.G. is supported by a Research Fellowship of
Jesus College, Cambridge, and G.L. is supported in part
by the NSF under grant PHY-91-57482.

\appendix
\section{Lightest BPS states}
\setcounter{equation}{0}

In this appendix we describe the lightest BPS states of type I' for
the first few values of the D0-brane number, and in the backgrounds
with $N=0,\ldots,8$ (table~3). These states were found by determining the
largest winding number for a physical BPS state with a given number of
D0-branes, {\it i.e.} the largest winding number that is consistent with
(\ref{condition}) and (\ref{IAcharges}). We use the convention that
all D0-branes are at $x^9=0$. 
\begin{table}[hbt]
\begin{center}
\begin{tabular}{|c||c|c|c|c|c|} \hline
   $N\backslash n_{I'}$ & 1/2 &1 & 3/2 & 2 & 5/2 \\
       \hline\hline
  0 & $0_{(\ss,\ii)+(\ii,\ss)}$ &
      $0_{(\aa,\ii)+(\ii,\aa)}$ &
      $0_{(\ss,\ii)+(\ii,\ss)}$ &
      $0_{(\aa,\ii)+(\ii,\aa)}$ &
      $0_{(\ss,\ii)+(\ii,\ss)}$ \\
     \hline
 1 &  $0_{(\ss,\ii)}$ & $0_{(\vv,\ii)}$ &
     $-1/2_{(\ss,\vv)}$ &
$-1/2_{(\vv,\vv)}$
         & $-3/4_{(\vv,\ss)}$ \\ \hline
 2 & $0_{(\ss,\ii)}$ & $0_{(\ii,\ii)}$ &
    $-1_{(\vv,\ss')+([\ss],[\ii])}$
    & $-1_{(\aa,\ii)+(\ii,\aa)}$ &
    $-3/2_{(\aa,\ss)+([\ss'],[\vv])}$ \\ \hline
 3 & $0_{(\ss,\ii)}$ & $-1/2_{(\ii,\vv)}$  &
    $-1_{(\ss',\ii)}$ &
    $-3/2_{(\vv,\vv)}$
    & $-2_{(\ss,\aa)}$  \\ \hline
 4 & $0_{(\ss,\ii)}$ &
     $-1_{(\aa,\ii)+(\ii,\aa)}$ &
    $-3/2_{(\ss',\vv)}$ &
    $-2_{(\aa,\ii)+(\ii,\aa)}$
    & $-5/2_{(\ss',\vv)}$ \\ \hline
 5 & $0_{(\ss,\ii)}$ & $-1_{(\vv,\ii)}$ &
     $-2_{(\ss',\aa)+([\ss'],\ii)}$ &
     $-5/2_{(\vv,\vv)}$
   & $-3_{(\ss,\ii)}$ \\ \hline
 6 & $0_{(\ss,\ii)}$ & $-1_{(\ii,\ii)}$ &
     $-2_{(\ss,\ii)}$  &
     $-3_{(\aa,\ii)+(\ii,\aa)}$
    & $-4_{(\vv,\ss')+([\ss],\ii)}$ \\ \hline
 7 & $0_{(\ss,\ii)}$ & $-3/2_{(\ii,\vv)}$  &
    $-5/2_{(\ss,\vv)}$ &
    $-7/2_{(\vv,\vv)}$
    & $-9/2_{([\ss'],\vv)}$ \\ \hline
 8 & $0_{\ii}$ & $-2_{\aa}$  & $-3_{\aa}$ &
    $-4_{\aa}$ & $-5_{\aa}$
\\ \hline
\end{tabular}
\caption{Winding number (assuming all D0-branes are at $x^9=0$)
and $SO(16-2N)\times SO(16+2N)$ charges of
lightest few charged type I' BPS states with $n_{I'}$ D0-branes.}
\end{center}
\end{table}

For $N=0$ and integer $n_{I'}$, there exist in addition massless
BPS-states with $N_L=1$ and $(P+2n_{I'}A)=0$; these are neutral states
(gravitons and photons). The representation $\ss$ corresponds to the
weight $(\pm 1/2)^L$ with an even number of $+$ signs, and the
representation $\ss'$ corresponds to an odd number of $+$ signs. For
$N=7$, the representations of $SO(2)\simeq U(1)$ are labeled by their
$U(1)$-charge $q\in\Zop/2$ modulo $2$, where $q_{ \ii}=0$,
$q_{\ss'}=1/2$, $q_{ \vv}=1$, and $q_{ \ss}=3/2$. Brackets denote
higher dimensional representations in the same conjugacy class of the
given representation.

Some of the states in table~3 can be obtained from the states with $N-1$
by moving a single D8-brane from $x^9=0$ to $x^9=2\pi$ as in
section~4. For example, the states $0_{(\aa,\ii)}$, $0_{(\vv,\ii)}$,
$0_{(\ii,\ii)}$, $-1/2_{(\ii,\vv)}$ and $-1_{(\ii,\aa)}$
at $n_{I'}=1$ and $N=0,1,2,3$ and $4$, respectively, are 
connected in this way. On the other hand the other state at $n_{I'}=1$
and $N=4$ ($-1_{(\aa,\ii)}$) does not arise in this way.

\section{D8-branes in the bulk}
\setcounter{equation}{0}

In this appendix we will derive the map between $(n_h,w_h)$ and
$(n_{I'},w_{I'})$, in the case where some D8-branes are in the bulk.
We shall consider specifically the situation where the
Wilson line of the heterotic theory is
\begin{equation}
A=(0^{8-N_1}, r^{N_1 -N_2}, (1/2)^{8+N_2})\,,
\label{wl}
\end{equation}
and $0 \leq r \leq 1/2$.

On the type I' side this corresponds to the configuration where
($8-N_1$) D8-branes are at $x^{9}=0$, ($N_1 -N_2$) D8-branes are at
$x^{9}=x^{9}(r)$, and ($8-N_2$) D8-branes are at $x^9 =2\pi$. (The
relation between $x^9(r)$ and $r$ will given below in (\ref{para}).)
The value of the background R-R 8-brane charge
is then $-N_1\mu_8$ in the region
$0 < x^9 < x^{9}(r)$, and $-N_2\mu_8$ in the region
$x^{9}(r) < x^{9} < 2\pi$.

Following \cite{polwit}, we can solve the equation of motion for the
type I' background, and we find that
\begin{eqnarray}
\frac{1}{R_{h}} & = & D M^{I'}_{winding}=\sqrt{\frac{8}{\alpha'_h}}
\frac{(a-b)/N_1+(c-a)/N_2}{[(a^2 -b^2 )/N_1+(c^2-a^2)/N_2]^{1/2}}\,,
\nonumber \\
D M^{I'}_{D0}(0) & =  & \sqrt{\frac{2}{\alpha'_h}}
\frac{b}{[(a^2 -b^2 )/N_1 +(c^2 -a^2)/N_2]^{1/2}}\,, \nonumber \\
r & = & \frac{1}{2} \frac{(a-b)/N_1}{(a-b)/N_1 + (c-a)/N_2}\,,
\label{para}
\end{eqnarray}
where we have defined
\begin{equation}
a=(B+x^{9}(r)N_{1})^{2/3}\,, \quad  b=B^{2/3}\,, \quad
c=(B+(N_1 -N_2)x^{9}(r) +2\pi x^{9}(r))^{2/3}\,.
\label{abc}
\end{equation}
As in section 3, we expect that a winding state in the heterotic
theory is mapped to a D0-brane together with some winding. However the
calculation is somewhat more elaborate as the connection between $r$
and $x^{9}(r)$ is rather complicated, as can be seen from equation
(\ref{para}).

We make the ansatz that
\be
\frac{R}{\alpha'_{h}} = \frac{1}{2} D M^{I'}_{D0}
+ D M^{I'}_{winding} \beta \,,
\label{def}
\ee
and solve for $\beta$, using the equations (\ref{para}).
After some algebra we find that
\be
\beta = \frac{N_1}{8} +\frac{(N_2 -N_1)}{8N_{2}^{2}} (c-a)^2 \,,
\label{bet}
\ee
which, in terms of the heterotic variables becomes
\begin{equation}
\beta =\frac{N_{1}}{8} +\frac{(r-1/2)^2}{2}(N_2 -N_1)\,.
\label{beth}
\end{equation}
Using equations (\ref{beth}, \ref{hetBPS}) the map between
$(n_h,w_h)$ and $(n_{I'}, w_{I'})$ is then given as
\begin{eqnarray}
w_{I'} & = & n_h -A\cdot P -w_h (1+N_2 /4 +(N_1 -N_2)r/2)\,, \nonumber
\\
n_{I'} & = & w_{h}/2\,.
\label{trans}
\end{eqnarray}
This formula is used in section~4 with $N_1=1$ and $N_2=0$.
It should be noted that these transformations can also be found simply by
assuming that massless $w_h=1$ states correspond to half
a D0-brane at $x^9=0$ without any winding.

\end{document}